\documentclass[aps,pra,twocolumn,showpacs]{revtex4-1}

\usepackage{amsmath}
\usepackage{amssymb}
\usepackage{bm}
\usepackage{textcomp}
\usepackage{graphicx}
\usepackage{bbold}

\begin{document}

\title{Long-range photon-mediated gate scheme between nuclear spin qubits in diamond}


\author{Adrian Auer}
\email{adrian.auer@uni-konstanz.de}
\author{Guido Burkard}
\affiliation{Department of Physics, University of Konstanz, D-78457 Konstanz, Germany}

\begin{abstract}
Defect centers in diamond are exceptional solid-state quantum systems that can have exceedingly long electron and nuclear spin coherence times. So far, single-qubit gates for the nitrogen nuclear spin, a two-qubit gate with a nitrogen-vacancy (NV) center electron spin, and entanglement between nearby nitrogen nuclear spins have been demonstrated. Here, we develop a scheme to implement a universal two-qubit gate between two distant nitrogen nuclear spins. Virtual excitation of an NV center that is embedded in an optical cavity can scatter a laser photon into the cavity mode; we show that this process depends on the nuclear spin state of the nitrogen atom. If two NV centers are simultaneously coupled to a common cavity mode and individually excited, virtual cavity photon exchange can mediate an effective interaction between the nuclear spin qubits, conditioned on the spin state of both nuclei, which implements a universal controlled-$\textit{Z}$ gate. We predict operation times below 100 nanoseconds, which is several orders of magnitude faster than the decoherence time of nuclear spin qubits in diamond.
\end{abstract}


\maketitle

\newcommand{\bra}[1]{\langle #1|}
\newcommand{\ket}[1]{|#1\rangle}
\newcommand{\braket}[2]{\langle #1|#2\rangle}
\newcommand{\nonum}{\nonumber \\}

\textit{Introduction.} Substantial experimental progress has been made in demonstrating the viability of nuclear spins coupled to nitrogen-vacancy (NV) centers in diamond as qubits. Compared with the NV electron spin, the nuclear spin offers for superior coherence properties, but so far, a scheme for the necessary two-qubit gates is lacking.
Candidate nuclear spins are the intrinsic nitrogen nuclear spin ($^{14}$N or $^{15}$N) \cite{PhysRevA.80.050302} or incidental proximal nuclear spins (e.g.~$^{13}$C) \cite{PhysRevLett.93.130501}. Decoherence times of $T_2^* \approx 5\,\textrm{ms}$ at room temperature have been measured \cite{Sar:2012}, and elementary single-qubit operations were implemented, including manipulation \cite{PhysRevA.80.050302,PhysRevB.81.035205,PhysRevLett.113.020506,sangtawesin15}, initialization \cite{PhysRevB.81.035205,PhysRevLett.102.057403,PhysRevA.80.050302,Neumann30072010,Robledo:2011} and high-fidelity single-shot readout \cite{PhysRevA.80.050302,Neumann30072010,Robledo:2011,PhysRevLett.110.060502,Waldherr:2014}. It was further demonstrated that the nitrogen nuclear spin can be a functioning part of a small quantum register \cite{Dutt01062007,Jiang09102009,Robledo:2011,Sar:2012,Pfaff:2013,Waldherr:2014}, or can act as a quantum memory to store and later retrieve the NV electron spin state \cite{Fuchs:2011}.  Nuclear spin entanglement has been studied both experimentally \cite{Neumann06062008,Pfaff:2013,Dolde:2013,Dolde:2014,H.:2014} and theoretically \cite{PhysRevA.72.052330,PhysRevLett.96.070504,PhysRevLett.107.150503,PhysRevX.4.031022}. However, a deterministic long-distance coupling scheme that does not utilize prior electron entanglement has not yet been demonstrated.  The coupling of nuclear spins is fundamentally required in the context of quantum information processing, e.g.~to perform universal quantum computation \cite{nielsen00}.

\begin{figure}[h!]
\includegraphics[width=0.95 \columnwidth]{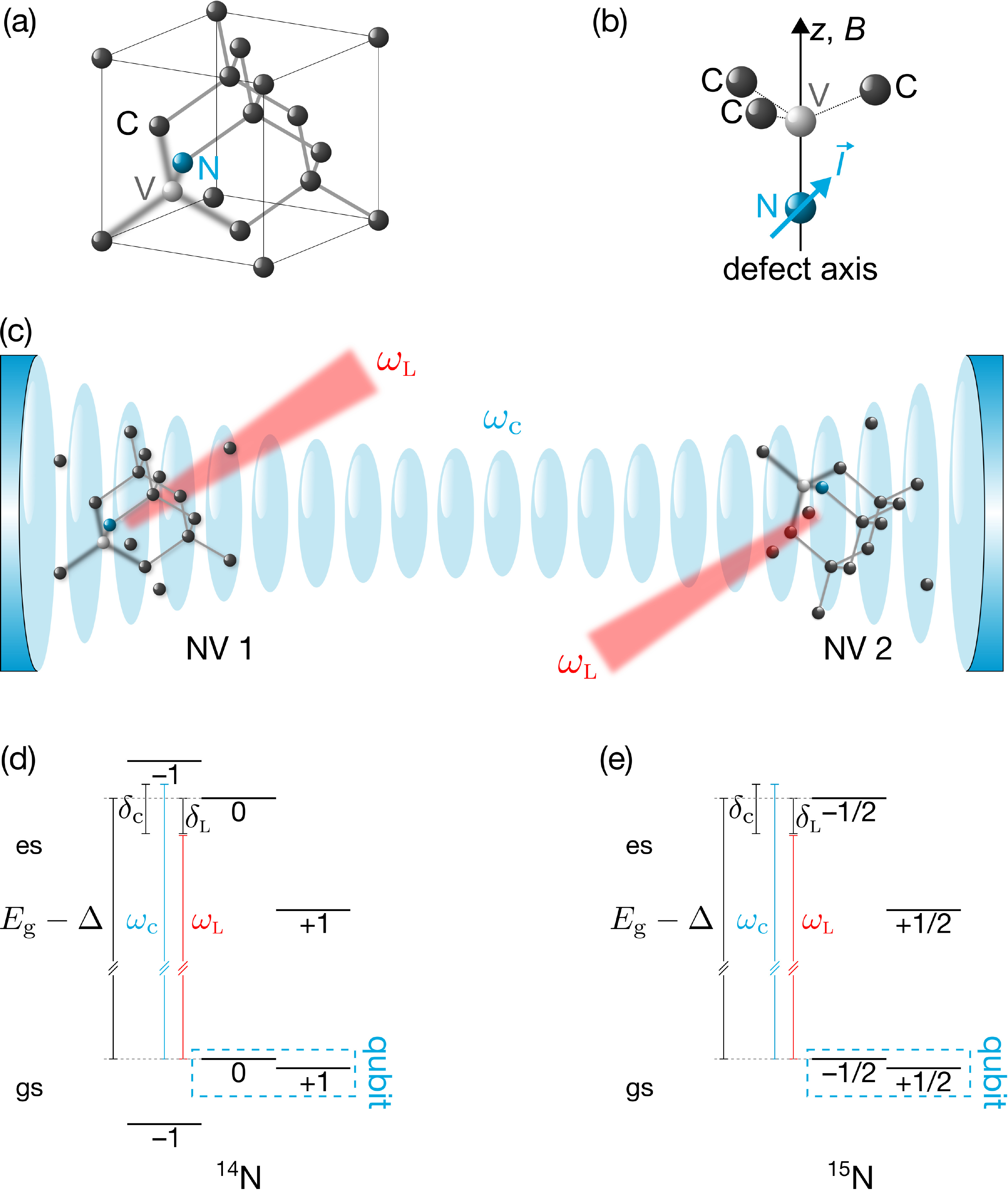}
\caption{\label{fig:NV_center}(a) NV center formed by a substitutional nitrogen (N) atom next to a vacancy (V) in the diamond 
lattice. (b) Magnetic field ($B$) direction along the defect axis ($z$) and N nuclear spin $\mathbf{I}$. (c) Interaction between two NV centers (NV 1 and NV 2) that are coupled to the same mode $\omega_\mathrm{c}$ of 
an optical cavity. The NV centers are excited by two lasers of frequency $\omega_\textsc{l}$. Scattering of a laser photon 
mediates an effective interaction between NV 1 and NV 2. (d) Hyperfine levels $m_I=0,\pm1$ of the $m_S=-1$ subspace 
for $^{14}$N in the ground (gs)~and excited (es) states, qubit states are indicated. 
$\delta_\textsc{l}$ and $\delta_\textrm{c}$  are the detunings of the laser frequency $\omega_\textsc{l}$ from the $m_I=0$ 
orbital transition energy $E_\textrm{g}-\Delta$ and from the cavity frequency $\omega_\textrm{c}$.
(e) Hyperfine levels $m_I=\pm1/2$ for $^{15}$N.}
\end{figure}

In this article, we 
develop and analyze a mechanism to optically generate a controlled quantum gate between two distant nitrogen nuclear spins (Fig.~\ref{fig:NV_center}). 
The coupling between the nuclear spins is achieved by exchanging virtual cavity photons among two NV centers. External laser photons incident on each NV center can be scattered into the cavity mode, or vice versa, by exciting electronic Raman-type transitions between the NV ground and excited state. We find that in the appropriate parameter regime, the scattering process depends on the nitrogen nuclear-spin state and can be completely suppressed for a specific nuclear spin configuration by properly tuning the laser frequency. This nuclear-spin dependent scattering mediates an effective interaction between two nitrogen nuclear spins. For a specific interaction time, a universal controlled-$Z$ ($\textsc{cz}$) gate is implemented, which is equivalent to \textsc{cnot} up to single qubit operations. A quantitative analysis of the proposed mechanism yields gate operation times below 100 nanoseconds, which is more than four orders of magnitude shorter than the decoherence time of several milliseconds for the nitrogen nuclear spin. While cavity-mediated coupling between NV \textit{electron} spins relies on the zero field splitting \cite{1402.6351}, the coupling of NV nuclear spins has its physical origin in the hyperfine interaction.

\textit{Model.} We start our analysis by describing a single NV center coupled to a single cavity mode and to an external laser field. The extension to two NV centers interacting with the same cavity mode, as required for the two-qubit gate, is straightforward and will be given later.
To model the combined system of a single NV center, an optical cavity and the external laser, we use the time-dependent Hamiltonian
\begin{equation}
\label{eq:Hamiltonian_single_NV}
H(t)
=
H_\textsc{nv} + H_\mathrm{c} + H_\textsc{l} (t) ,
\end{equation}
where $H_\textsc{nv} = H_\mathrm{e} + H_\mathrm{n} + H_\mathrm{hf}$ describes the electron (e) and nuclear (n) spin systems coupled through hyperfine (hf) interactions, $H_\mathrm{c}$ the coupling to the cavity, and $H_\textsc{l} (t)$ the interaction with the laser field. 
In the presence of an external magnetic field $\mathbf{B} = B \mathbf{e}_z$ along the defect symmetry axis ($z$ axis), the electron spin ($\mathbf{S}$) and nuclear spin ($\mathbf{I}$) Hamiltonians are given by ($\hbar = 1$)\cite{1367-2630-13-2-025019,Doherty20131,1402.6351}
\begin{align}
\label{eq:Hamiltonian_electronic_spin}
H_\mathrm{e}
&=
\gamma_\mathrm{e} B S_z + D S_z^2  - \frac{1}{2} \Delta S_z^2 \tau_z + \frac{1}{2} E_\mathrm{g} \tau_z,
\\
H_\mathrm{n}
&=
- \gamma_\mathrm{n} B I_z + Q I_z^2.
\end{align}
Here, $\gamma_\mathrm{e}/2\pi = 2.803$ MHz/G is the electron gyromagnetic ratio, $E_\mathrm{g} = 1.945$ eV is the energy gap between ground and excited state, and $D=(D_\mathrm{gs} + D_\mathrm{es})/2$ and $\Delta = D_\mathrm{gs} - D_\mathrm{es}$ with the zero-field spin splittings of the ground ($D_\mathrm{gs}/2\pi = 2.88$ GHz) and excited state ($D_\mathrm{es}/2\pi= 1.42$ GHz). The nuclear gyromagnetic ratio is denoted $\gamma_\mathrm{n}$ and Q is the nuclear electric quadrupole coupling (see Tab.~\ref{tab:values}). To describe the orbital degree of freedom, we use Pauli matrices $\tau_i$ $(i=x,y,z)$, and choose the ground and excited states as $\tau_z$ eigenvectors with eigenvalues $-1$ and $+1$, respectively. We only consider the lower orbital branch of the excited state doublet ($E_y$). This is justified by naturally occurring strain fields of 10 GHz and more \cite{doi:10.1146/annurev-conmatphys-030212-184238}, which split the excited state into two well-separated orbital branches.

Hyperfine interaction in the excited state is modeled by a diagonal hyperfine tensor, which has the same form as in the ground state \cite{PhysRevB.80.241204,Ivady2015}. However, since the electron density at the nitrogen site is larger in the excited state \cite{PhysRevB.77.155206}, the hyperfine interaction is about 20 times stronger compared to the ground state according to measurements under ambient conditions \cite{PhysRevLett.101.117601,PhysRevB.81.035205}. The difference $\delta A$ between the hyperfine coupling in the ground and the excited state forms the basis of the nuclear-spin dependent light scattering effect which we predict.
Working at magnetic field strengths away from the ground and excited state level anticrossings, electron-nuclear spin flip-flop processes are energetically suppressed. Therefore, we neglect the transverse part of the hyperfine tensor and only include the longitudinal coupling. Denoting the hyperfine coupling strengths by $A_\mathrm{gs}$ and $A_\mathrm{es}$ for the ground and excited state (Tab.~\ref{tab:values}), we arrive at
\begin{equation}
H_\mathrm{hf}
=
A S_z I_z +\frac{1}{2} \delta A \: \tau_z S_z I_z,
\end{equation}
where $A = (A_\mathrm{es} + A_\mathrm{gs})/2$ and $\delta A = A_\mathrm{es} - A_\mathrm{gs}$.

\begin{table}
\caption{\label{tab:values}Relevant nuclear-spin parameters for the NV center.}
\begin{ruledtabular}
\begin{tabular}{lcc}
Parameter & $^{14}$N & $^{15}$N \\
\hline
Nuclear spin $I$ & 1 & 1/2\\
$\gamma_\mathrm{n}/2\pi$ & 0.308 kHz/G \cite{NMR2001} & -0.432 kHz/G \cite{NMR2001}\\
$Q/2\pi$ & -5 MHz \cite{PhysRevB.79.075203,PhysRevA.80.050302,PhysRevB.81.035205} & 0 \\
$A_\mathrm{gs}/2\pi$ & -2.2 MHz \cite{PhysRevB.79.075203,PhysRevA.80.050302,PhysRevB.81.035205,PhysRevLett.113.020506} & 3.0 MHz \cite{PhysRevLett.101.117601,PhysRevB.79.075203}\\
$A_\mathrm{es}/2\pi$ & 40 MHz \cite{PhysRevB.81.035205} & 61 MHz \cite{PhysRevLett.101.117601}\\
\end{tabular}
\end{ruledtabular}
\end{table}

We consider the NV electronic orbital transition between the ground and excited state to be coupled to a single mode of the optical cavity, which, in the rotating-wave approximation,  is described by 
$
H_\mathrm{c}
=
\omega_\mathrm{c} a^\dagger a + g (\tau_+ a + \tau_- a^\dagger)
$,
where $\omega_\mathrm{c}$ is the cavity frequency, $a^{(\dagger)}$ is cavity-photon annihilation (creation) operator, $g$ the coupling strength (which be assumed real), and $\tau_\pm = (\tau_x \pm i \tau_y) / 2$.
The external laser is described by a classical field of frequency $\omega_\textsc{l}$ that excites electronic orbital transitions between states having the same spin projections $m_S$ and $m_I$,
$
H_\textsc{l} (t)
=
\Omega e^{- i \omega_\textsc{l} t} \tau_+ + \Omega^* e^{ i \omega_\textsc{l} t} \tau_-
$.
Here, $\Omega$ is the complex Rabi frequency that depends on the phase of the laser field.
The Hamiltonian $H(t)$ can be made time-independent by transforming into a rotating frame, $H'=e^{i \xi t} H(t) e^{- i \xi t} - \xi$ with $\xi = \omega_\textsc{l} (a^\dagger a + \tau_z/2)$, and we obtain
$
H'
=
H_\mathrm{e}' + H_\mathrm{n} + H_\mathrm{hf} + H_\mathrm{c}' + H_\textsc{l}'
$.
The transformed part $H_\mathrm{e}'$ of the electronic Hamiltonian is obtained by replacing $E_\mathrm{g}$ with the detuning $\tilde{\delta}_\textsc{l} = E_\mathrm{g} - \omega_\textsc{l}$ in $H_\mathrm{e}$.
In the Hamiltonian $H_\mathrm{c}$, the transformation causes a shift of the cavity frequency to $\delta_\mathrm{c} = \omega_\mathrm{c} - \omega_\textsc{l}$, which is the detuning of the laser from the cavity mode. The laser Hamiltonian $H_\textsc{l} (t)$ becomes time-independent, $H_\textsc{l}' = \Omega \tau_+ + \Omega^* \tau_-$.

\textit{Nuclear-spin dependent scattering.} Virtually exciting the NV center by the external laser field can finally lead to an excitation of the cavity mode through the coupling $g$. We describe this process by using quasi-degenerate perturbation theory in terms of a Schrieffer-Wolff (SW) transformation \cite{PhysRev.149.491,Winkler:2003} to eliminate the intermediate virtual transition to the excited state, and obtain a model that effectively describes the scattering of a laser photon into the cavity mode, and vice versa, that particularly depends on the nitrogen nuclear spin projection $m_I$. It is exactly this spin-dependent scattering that eventually enables a conditional two-qubit quantum gate.

To implement the SW transformation, we construct an anti-Hermitian operator $S$ such that  $[S, H_0] = V$ (see Appendix), where the part $
H_0
=
H_\mathrm{e}' + H_\mathrm{n} + H_\mathrm{hf} + \delta_\mathrm{c} a^\dagger a$
only acts on the ground and excited state manifold, respectively, and $V$ describes transitions between these two Hilbert subspaces.
In the transformed Hamiltonian $\widetilde{H} = e^{-S} H' e^{S}$, we keep the lowest order in the interaction $V$ and continue with the effective Hamiltonian $\widetilde{H} \approx H_0 + [V,S]/2$.
The effective ground-state Hamiltonian becomes (see Appendix)
\begin{align}
\label{eq:effective_Hamiltonian_ground_state}
&\widetilde{H}^\mathrm{(gs)}
=
- \left( A_\mathrm{gs} + \gamma_\mathrm{n} B \right) I_z + Q I_z^2 + \delta_\mathrm{c} a^\dagger a 
 \\
&+ \frac{1}{2} \Big[
g \Omega \left(
\left(
\delta A \: I_z - \delta_\textsc{l}
\right)^{-1}
+
\left(
\delta A \: I_z + \delta_\mathrm{c} - \delta_\textsc{l}
\right)^{-1}
\right)
a^\dagger
+ \mathrm{h.c.}
\Big].
\nonumber
\end{align}
Here, we restrict our consideration to the $m_S = -1$ subspace, and define the detuning $\delta_\textsc{l} = \tilde{\delta}_\textsc{l} - \Delta$ of the laser frequency from the $m_I=0$ orbital transition (Fig.~\ref{fig:NV_center}). We omit all constant terms and neglect small energy shifts proportional to $g^2$ and $|\Omega|^2$.

On the basis of previous experimental work \cite{Fuchs:2011,Pfaff:2013,PhysRevLett.113.020506} using the ${}^{14}$N nuclear spin as a qubit, we choose the nuclear spin sublevels $\ket{m_I = 1}= \ket{1}$ and $\ket{m_I = 0} = \ket{0}$ as the computational basis. We can neglect the $m_I = -1$ state because the transition frequency between these two levels is well separated from other transitions \cite{PhysRevLett.113.020506}. From Eq.~(\ref{eq:effective_Hamiltonian_ground_state}), one can see that the effective coupling of the NV center to the cavity via the virtual laser excitation depends on the spin state of nitrogen nucleus and can, e.g., be completely suppressed for one of the two spin states. This is the case if the laser frequency is chosen such that e.g.~$\delta_\textsc{l} = \delta_\mathrm{c}/2$, where only scattering from the $m_I = 1$ state is possible. By using $I_z = \ket{1} \bra{1}$ and $\mathbb{1} = \ket{1} \bra{1} + \ket{0} \bra{0}$, we find the qubit Hamiltonian
\begin{align}
\widetilde{H}^\mathrm{(qubit)}
&=
\left( Q - A_\mathrm{gs} - \gamma_\mathrm{n} B \right) \ket{1} \bra{1} + \delta_\mathrm{c} a^\dagger a
\nonum
&+ g' \ket{1} \bra{1} a^\dagger + (g')^{*} \ket{1} \bra{1} a,
\label{eq:Hamiltonian_mI1_scattering}
\end{align}
with an effective coupling strength
\begin{equation}
\label{eq:effective_single_qubit_coupling}
g'
=
g \Omega \frac{\delta A}{\delta A^2 - \left( \frac{\delta_\mathrm{c}}{2} \right)^2}.
\end{equation}
Scattering only from the $m_I = 0$ state is possible for $\delta_\textsc{l} = \delta A + \delta_\mathrm{c}/2$ occurring with the same effective coupling strength $g'$ [Eq.~(\ref{eq:effective_single_qubit_coupling})]; however, we concentrate on $m_I = 1$ scattering in the following.

\begin{figure}
\includegraphics[width=\columnwidth]{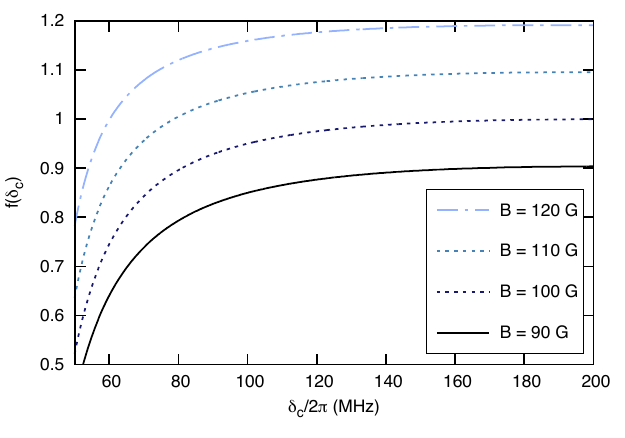}
\caption{\label{fig:comparison_gEff}Ratio $f(\delta_\mathrm{c})=\tilde{g}/g'$ of coupling strengths with ($\tilde{g}$) and without ($g'$) spin-spin interaction in the excited state. Magnetic field strengths $B$ are chosen such that $m_I$ is a good quantum number in the ground and excited state.}
\end{figure}

\textit{Spin-spin interactions.} To understand the scattering mechanism of a laser photon into the cavity mode qualitatively, we so far neglected spin-mixing terms in the lower branch of the excited state doublet \cite{PhysRevLett.101.117601,PhysRevLett.102.195506,1367-2630-13-2-025019,1367-2630-13-2-025025,Bassett12092014}. However, to make quantitative predictions of the effective scattering process, we take into account the fine structure of the excited state manifold.
So far, electronic spin-spin interactions were only incorporated by the zero-field splittings $D_\mathrm{gs.}$ and $D_\mathrm{es}$. In the limit of high strain considered here, the two branches of the excited-state orbital doublet split and anticrossings in the lower branch mix spin states with different quantum numbers $m_S$.
The Hamiltonian describing the spin mixing is \cite{Doherty20131,Bassett12092014}
\begin{equation}
H_\mathrm{s}
=
\frac{1}{2} (1 + \tau_z)
\left(
\frac{\Delta_1}{2} \left( S_x^2 - S_y^2 \right) - \frac{\Delta_2}{\sqrt{2} }\left(S_x S_z + S_z S_x \right)
\right),
\end{equation}
where transitions between the excited state orbitals have been neglected due to the high strain, and the fine structure parameters are given by $\Delta_1/2\pi = 1.54$ GHz and $\Delta_2/2\pi = 0.154$ GHz \cite{Bassett12092014}. The effective coupling strength $\tilde{g}$ analogous to Eq.~(\ref{eq:effective_single_qubit_coupling}) can be obtained by adding $H_\mathrm{s}$ to the bare NV Hamiltonian $H_\textsc{nv}$, and then performing the SW transformation. In doing so, we assume the cavity to be populated by at most one photon, and only if the NV center is in the ground state. In the excited state, we need to include all spin states $m_S=0,\pm1$. The effective ground state Hamiltonian in the case of $m_I = 1$ scattering has the same form as given in Eq.~(\ref{eq:Hamiltonian_mI1_scattering}) with a different coupling strength $\tilde{g}
=
g' f(\delta_\mathrm{c})
$.
The detuning-dependent part $f(\delta_\mathrm{c})$ is plotted in Fig.~\ref{fig:comparison_gEff}. 

\begin{figure}
\includegraphics[width=\columnwidth]{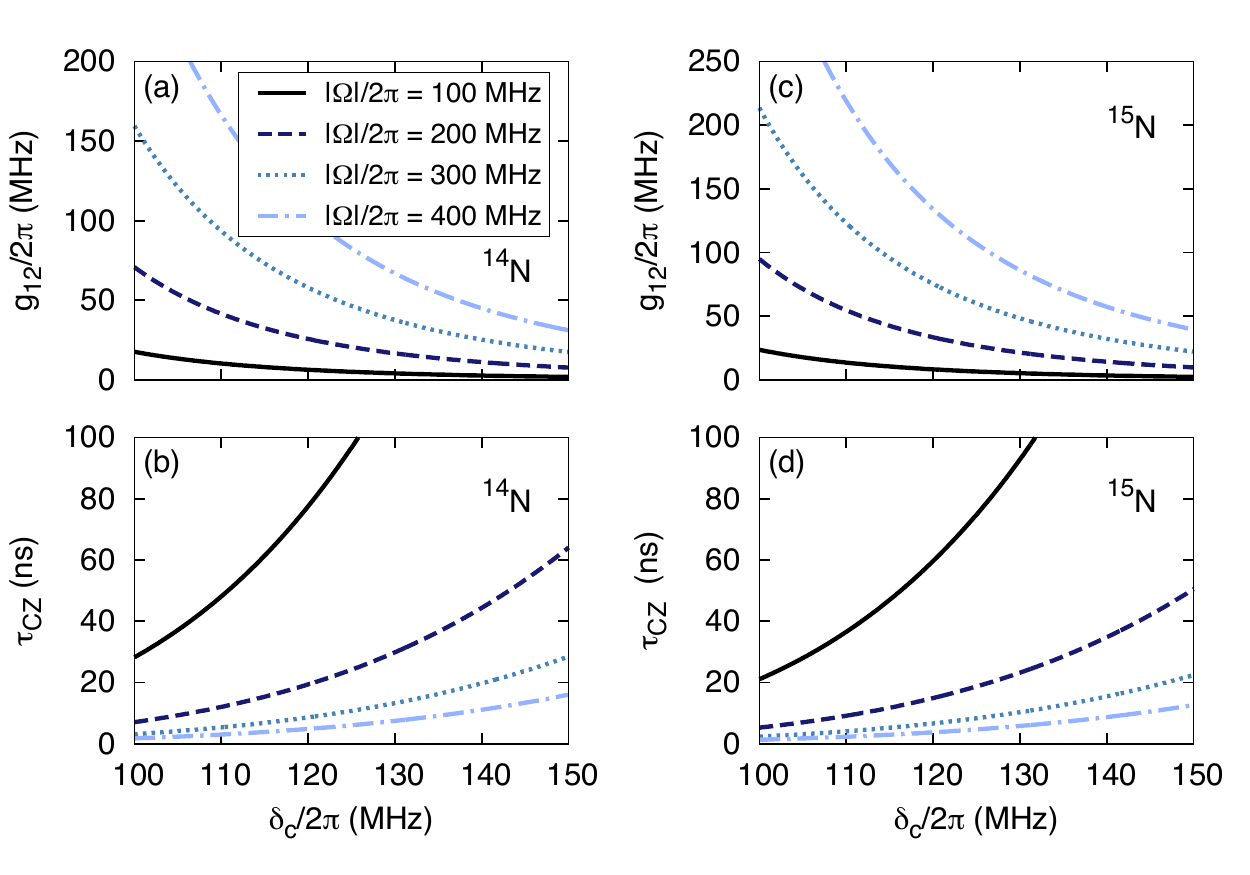}
\caption{\label{fig:g12}(a) Effective two-qubit coupling strength $g_{12}$ as a function of $\delta_\mathrm{c}$ for different values of the laser Rabi frequency $\Omega_1=\Omega_2\equiv\Omega$ (see legend, valid for all figures) and $g_1/2\pi =g_2/2\pi = 50$ MHz for ${}^{14}$N nuclear spins. (b) Time $\tau_\textsc{cz}$ to generate a \textsc{cz} gate between the two nuclear spins as a function of $\delta_\mathrm{c}$ for $g_1/2\pi =g_2/2\pi = 50$ MHz. (c) and (d) equivalent for
 ${}^{15}$N nuclear spins. All calculations performed at $B=120$ G.}
\end{figure}

\textit{Controlled quantum gate.} For the two-qubit gate, we consider two NV centers ($i=1,2$) coupled to the same cavity mode and each individually driven by a laser of frequency $\omega_\textsc{l}$ [Fig.~\ref{fig:NV_center} (c)].
In the following, we keep only the lowest order of the interaction parts, and consider $m_I = 1$ scattering on both NV centers.
Furthermore, we assume detunings $\delta_\textsc{l}$ and $\delta_\textrm{c}$ such that the cavity is excited only virtually, which, in turn, leads to an effective interaction between the two NV centers. To describe this interaction, we apply a second SW transformation to $ \widetilde{H}_2^\mathrm{(gs)} =
\delta_\mathrm{c} a^\dagger a + \sum_{i = 1}^2  \left( Q - A_\mathrm{gs} - \gamma_\mathrm{n} B \right) \ket{1}_i \bra{1}
+
 (\tilde{g}_i \ket{1}_i \bra{1} a^\dagger + \textrm{h.c.})$  to eliminate the cavity mode
by choosing
$
S
=
- \sum_{i = 1}^2 (\tilde{g}_i/\delta_\mathrm{c} \ket{1}_i \bra{1} a^\dagger - \mathrm{h.c.})$,
which leads to an effective Hamiltonian $H_\mathrm{eff}= e^{-S}\widetilde{H}_2^\mathrm{(gs)} e^S$, where again only the lowest order contribution of the off-diagonal elements is kept (see Appendix).
$H_\mathrm{eff}$ comprises single-qubit terms
$H_\mathrm{eff}^{(i)} 
=
( Q - A_\mathrm{gs} - \gamma_\mathrm{n} B - |\tilde{g}_i|^2/\delta_\mathrm{c}) \ket{1}_i \bra{1},
$
and a two-qubit interaction term,
\begin{equation}
H_\mathrm{int}
=
- g_{12} \ket{11} \bra{11}.
\end{equation}
Here, $\ket{11} = \ket{1}_1 \ket{1}_2$ is the nuclear spin state of both NV centers 1 and 2, and the effective two-qubit coupling strength $g_{12}$ is found to be
\begin{equation}
g_{12}
=
2 \frac{|\tilde{g}_1||\tilde{g}_2|}{\delta_\mathrm{c}} \cos (\phi_1 - \phi_2),
\end{equation}
where $\phi_i$ denotes the phase of the $i$th laser field, $\Omega_i = |\Omega_i|e^{i\phi_i}$. Quantitative predictions of $g_{12}$ are plotted in Fig.~\ref{fig:g12} (a).

Since $[H_\mathrm{eff}^{(i)},H_\mathrm{int}] = 0$, the time evolution $U$ generated by the Hamiltonian $H_\mathrm{eff}$ can be written as
\begin{equation}
\label{eq:time_evolution}
U(t)
=
e^{-i H_\mathrm{eff} t}
=
\left(
U_1 (t) \otimes U_2 (t)
\right)
U_{12} (t),
\end{equation}
where $U_i (t)$ is a single-qubit rotation of nuclear spin $i$ and $U_{12} (t)$ describes a two-qubit operation generated by the interaction part $H_\mathrm{int}$. In Eq.~(\ref{eq:time_evolution}), the time evolution of the cavity field has been omitted, since the nuclear spin degree of freedom has been decoupled from the cavity field by the above transformation. In the following, we only concentrate on the two-qubit interaction part, and disregard single-qubit rotations since they can be undone afterwards, e.g., by off-resonant excitation of the ground-state electronic spin transition, thereby implementing a phase gate on the N nuclear spin \cite{PhysRevLett.113.020506}, or direct driving of the nuclear spin transitions \cite{PhysRevA.89.052317}.

For an operation time of $\tau_\textsc{cz} = \pi/ g_{12}$, a \textsc{cz} gate is implemented on the two nuclear spin qubits,
\begin{equation}
U_{12} (\tau_\textsc{cz})
=
\ket{00} \bra{00} + \ket{01} \bra{01} + \ket{10} \bra{10} - \ket{11} \bra{11},
\end{equation}
from which \textsc{cnot} can be created using additional Hadamard gates \cite{nielsen00}.
In Fig.~\ref{fig:g12} (b), values of $\tau_\textsc{cz}$ are shown for different Rabi frequencies $\Omega$. As the main result of our paper, we find fast operation times clearly below 100 ns. In our calculations, we assumed large detunings $|\delta_\textsc{l} - \delta_\mathrm{c}| \gtrsim g$ and $|\delta_\textsc{l}| \gtrsim |\Omega|$ to justify the effective model used.

\textit{Conclusions.} Nitrogen nuclear spins in diamond have proved to be highly promising candidates to physically realize qubits. We have presented a theoretical proposal for the implementation of a controlled optical cavity-mediated quantum gate between two nitrogen nuclear spin qubits intrinsic to NV centers in diamond. Gate operation can be achieved within 100 nanoseconds or less, which is more than four orders of magnitude below the nuclear-spin decoherence time. Assuming $\tau_\textsc{cz} \approx 20$ ns [see Fig.~\ref{fig:g12} (b)], the cavity loss rate $\kappa$ must not exceed values of $1/\tau_\textsc{cz} \approx 50$ MHz, which requires $Q$ factors of $10^6$--$10^7$ \cite{1402.6351}. High-$Q$ silica microsphere cavities can reach such values \cite{doi:10.1021/nl061342r} and progress towards photonic crystal cavities in bulk diamond exceeding $Q$-factors of $10^5$ has been recently achieved \cite{Burek:2014}.

In addition to the presented findings, an equivalent analysis for the ${}^{15}$N nuclear spin with $I=1/2$ show that the proposed scheme also works for this isotope if the computational basis is chosen as $\ket{1}=\ket{m_I=+1/2}$ and $\ket{0}=\ket{m_I=-1/2}$ [Fig.~\ref{fig:NV_center} (e)]. We find the same effective scattering rate $g'$ [Eq.~(\ref{eq:effective_single_qubit_coupling})] for $m_I=\pm1/2$ scattering for laser detunings $\delta_\textsc{l} = (\delta_\textrm{c} \mp \delta A)/2$. Including spin-spin interactions, the effective two-qubit coupling strength $g_{12}$ and the gate time $\tau_\textsc{cz}$ show qualitatively the same behavior as for the ${}^{14}$N nuclear spin, and are depicted in Figs.~\ref{fig:g12} (c) and (d).
During the fast electronic excitation cycles, the nuclear spins are subject to a time-varying hyperfine interaction. However, using a spin-fluctuator model, it has been shown that nuclear spin state will be unaffected and coherence can be preserved \cite{PhysRevLett.100.073001}. 
Together with elementary and experimentally demonstrated single-qubit operations, the realization of a universal \textsc{cz} gate makes the nitrogen nuclear spin valuable  for quantum computation in addition to its remarkable quality as a quantum memory \cite{Fuchs:2011}.

\textit{Acknowledgements.} We acknowledge funding from the DFG within SFB 767 and from the BMBF under the program Q.com-HL.

\appendix


\section{Schrieffer-Wolff transformation to eliminate excited state}
We separate the Hamiltonian $H'$ into a block-diagonal part $H_0$ that only acts within the ground and the excited state manifold, respectively, and an off-diagonal part $V$ that connects these two manifolds,
\begin{equation}
H' = H_0 + V.
\end{equation}
To implement the Schrieffer-Wolff (SW) transformation \cite{PhysRev.149.491,Winkler:2003}, we construct a unitary transformation $\exp(-S)$ with some anti-Hermitian $S$ to obtain a new Hamiltonian $\widetilde{H}$,
\begin{equation}
\widetilde{H}
=
e^{-S}H'e^S,
\end{equation}
that contains no matrix elements that connect the ground and the excited states up to a desired order in $V$. If we choose the anti-Hermitian operator $S$ in such a way that 
\begin{equation}
\label{eq:condition}
[S, H_0] = V
\end{equation}
holds, the leading order in $V$ cancels. If we keep the lowest order in $V$, the Hamiltonian $\widetilde{H}$ is approximately given by
\begin{equation}
\widetilde{H}
\approx
H_0 + \frac{1}{2} [V,S].
\end{equation}

The block-diagonal part $H_0$ of $H'$ is given by
\begin{equation}
H_0
=
H_\mathrm{e}' + H_\mathrm{n} + H_\mathrm{hf} + \delta_\mathrm{c} a^\dagger a,
\end{equation}
and the interaction terms are
\begin{equation}
V
=
g (\tau_+ a + \tau_- a^\dagger) + \Omega \tau_+ + \Omega^* \tau_-.
\end{equation}
From the condition in Eq.~(\ref{eq:condition}), we find
\begin{align}
S
&=
\Omega \left( \Delta S_z^2 - \Delta_\mathrm{hf} S_z I_z - \tilde{\delta}_\textsc{l} \right)^{-1} \tau_+ - \mathrm{h.c.}
\nonumber \\
&+ g \left( \Delta S_z^2 - \Delta_\mathrm{hf} S_z I_z + \delta_\mathrm{c} - \tilde{\delta}_\textsc{l} \right)^{-1} \tau_+ a - \mathrm{h.c.},
\label{eq:S_single_qubit}
\end{align}
and the effective Hamiltonian for the decoupled ground state manifold becomes
\begin{align}
&\widetilde{H}^\mathrm{(gs)}
=
- \left( A_\mathrm{gs} + \gamma_\mathrm{n} B \right) I_z + Q I_z^2 + \delta_\mathrm{c} a^\dagger a 
\nonumber \\
&+ \frac{1}{2} \Big[
g \Omega \left(
\left(
\Delta_\mathrm{hf} I_z - \delta_\textsc{l}
\right)^{-1}
+
\left(
\Delta_\mathrm{hf} I_z + \delta_\mathrm{c} - \delta_\textsc{l}
\right)^{-1}
\right)
a^\dagger
\nonumber \\
&
+ \mathrm{h.c.}
\Big].
\label{eq:effective_Hamiltonian_ground_state}
\end{align}
Here, we restrict our consideration to the $m_S = -1$ subspace, and define the detuning $\delta_\textsc{l} = \tilde{\delta}_\textsc{l} - \Delta$ of the laser frequency from the $m_I=0$ orbital transition. We omit all constant terms and neglect small energy shifts proportional to $g^2$ and $|\Omega|^2$.

\section{SW transformation to eliminate virtual photon}

We start from a Hamiltonian $\widetilde{H}_2^{(\textrm{gs})}$ that describes two NV centers ($i=1,2$) coupled to a common cavity mode and each driven by a laser of frequency $\omega_\textsc{l}$,
\begin{align}
\widetilde{H}_2^\mathrm{(gs)}
&=
\delta_\mathrm{c} a^\dagger a + \sum \limits_{i = 1}^2  \left( Q - A_\mathrm{gs} - \gamma_\mathrm{n} B \right) \ket{1}_i \bra{1}
\nonum
&+
 \tilde{g}_i \ket{1}_i \bra{1} a^\dagger + \tilde{g}_i^* \ket{1}_i \bra{1} a,
\label{eq:Hamiltonian_two_qubits}
\end{align}
where we consider $m_I=1$ scattering on both NV centers and assume detunings $\delta_\textsc{l}$ and $\delta_\textrm{c}$ such that the cavity is excited only virtually. The effective coupling strength $\tilde{g}_i$ is given by
\begin{equation}
\tilde{g}_i
=
g_i' f(\delta_\textrm{c})
=
g_i \Omega_i \frac{\delta A}{\delta A^2 - \left( \frac{\delta_\textrm{c}}{2} \right)^2}f(\delta_\textrm{c}),
\end{equation}
where $g_i$ is the coupling strength of NV center $i$ to the cavity and $\Omega_i$ is the Rabi frequency of the $i$th laser field.

To derive an effective interaction between the two nuclear spin qubits, we apply a second SW transformation to eliminate the cavity mode, i.e.~to decouple the subspaces containing zero and one cavity photon, by choosing
\begin{equation}
S
=
- \sum_{i = 1}^2 \left(\frac{\tilde{g}_i}{\delta_\mathrm{c}} \ket{1}_i \bra{1} a^\dagger - \mathrm{h.c.}\right).
\end{equation}
We obtain an effective Hamiltonian through the unitary transformation
\begin{equation}
\label{eq:effective_H}
H_\textrm{eff}
=
e^{-S}\widetilde{H}_2^\mathrm{(gs)}e^S\approx
\sum \limits_{i=1}^2 H_\textrm{eff}^{(i)} + H_\textrm{int} +\delta_\textrm{c} a^\dagger a,
\end{equation}
where we also keep terms up to the lowest order in the off-diagonal matrix elements. The Hamiltonian $H_\textrm{eff}$ contains terms that only act on a single nuclear spin $i$,
\begin{equation}
H_\mathrm{eff}^{(i)} 
=
\left( Q - A_\mathrm{gs} - \gamma_\mathrm{n} B - \frac{|\tilde{g}_i|^2}{\delta_\mathrm{c}}\right) \ket{1}_i \bra{1},
\end{equation}
and an interaction part $H_\textrm{int}$ that couples the two nuclear spin qubits,
\begin{equation}
H_\mathrm{int}
=
- g_{12} \ket{11} \bra{11}.
\end{equation}
The last term in Eq.~(\ref{eq:effective_H}) is zero in the considered subspace that contains no photons.



%

\end{document}